\documentclass[pra,twocolumn,showpacs]{revtex4-1}

\usepackage{graphicx} 

\begin{document}

\title{Continuous-variable qubit on an optical transverse mode}
\author{Makoto Takeuchi}
 \email{takeuchi@phys.c.u-tokyo.ac.jp} 
\affiliation{Department of Basic Sciences, Graduate School of Arts and Sciences, The University of Tokyo, 153-8902 Japan}

\author{Takahiro Kuga}
\affiliation{Department of Basic Sciences, Graduate School of Arts and Sciences, The University of Tokyo, 153-8902 Japan}

\date{October 7, 2013}

\begin{abstract}
Continuous-variable (CV) qubits
can be created
 on an optical longitudinal mode
in which quantum information is encoded
 by the superposition of even and odd Schr\"{o}dinger's cat states with quadrature amplitude.
Based on the analogous features of paraxial optics and quantum mechanics,
 we propose a system to generate and detect CV qubits on an optical transverse mode.
As a proof-of-principle experiment, we generate six CV qubit states
 and observe their probability distributions in position and momentum space.
This enabled us to prepare a non-Gaussian
initial state for CV quantum computing.
Other potential applications of the CV qubit
include adiabatic control of a beam profile,
 phase shift keying on transverse modes,
 and quantum cryptography using CV qubit states.
\end{abstract}


\maketitle

\section{Introduction} \label{Sec:Intro}

Continuous-variable (CV) quantum computing
 is a potentially useful method for performing several types of task more efficiently
 than is possible with standard (discrete-variable) quantum computing \cite{Lloyd1999}.
To perform CV quantum computing,
it must be possible to apply an arbitrary time evolution $\exp(-i\hat{G})$
to the continuous variables ($\hat{X}$ and $\hat{P}$) at will.
Although in practice it is difficult to apply interactions $\hat{G}$
that are represented by greater than second-order polynomials of $\hat{X}$ and $\hat{P}$,
 if a non-Gaussian state can somehow be prepared \cite{Ralph2003},
 the quadratic interaction $\hat{G}^{(2)}$ will be suitable for performing CV quantum computing.
Photon subtraction from a squeezed vacuum state
 is a good way to prepare non-Gaussian states in optical longitudinal modes
 \cite{Neergaard-Nielsen2006,Wakui2007,Takahashi2008}, and it
 has been demonstrated that, by combining the photon subtraction and displacement operations, a class of non-Gaussian states can be prepared
 within a given system \cite{Neergaard-Nielsen2010}.
Because its states can ideally be represented using the qubits
 $C_\uparrow|\!\!\uparrow_\omega\rangle+C_\downarrow|\!\!\downarrow_\omega\rangle$,
 where $C_{\uparrow,\downarrow}$ are complex numbers and
 $|\!\!\uparrow_\omega\rangle$ and $|\!\!\downarrow_\omega\rangle$
 are even and odd Schr\"{o}dinger's cat states,respectively, on the longitudinal mode, this class is called the ``continuous-variable qubit."
The even and odd cats represent superpositions of two macroscopic states,
 where even and odd denote, respectively, whether the interference is constructive or destructive.
In the case of optical longitudinal modes, the macroscopic state represents coherent (classical) light. In practice, however, CV qubits generated by photon subtraction will not be ideal;
rather, they will be probabilistically generated states that are mixed to a certain extent with vacuum fluctuation.
As a result, the purity around the $|\!\!\downarrow_\omega\rangle$
 state will be far from unity (as shown Figure 11 in Ref. \cite{Neergaard-Nielsen2011}).
In addition, as the success rate in generating the even cat state is not very high,
 it must be substituted for with a squeezed vacuum.
Thus, more innovations will be necessary in order to achieve
 CV quantum computing on optical longitudinal modes.

In addition to longitudinal modes, light also has transverse modes
and, as analogies can be found between paraxial wave optics and quantum mechanics, CV quantum computing on optical transverse modes
may be possible \cite{Man'ko2001, Dragoman2002}.
In this work, we will assume that light is always in a coherent state
in its longitudinal mode, either in free space or in linear optical elements such as lenses and beam splitters.
Based on the formulation in Ref. \cite{Nienhuis1993}, we can quantize the transverse mode as
\begin{eqnarray}
\left[\begin{array}{c}
\hat{a}_x \\
\hat{a}_y \\
\end{array}\right]
\equiv \frac{\sqrt{2}}{w_0}
\left[\begin{array}{c}
\hat{x} \\
\hat{y} \\
\end{array}\right]
+i\frac{w_0}{\sqrt{2}\hbar}
\left[\begin{array}{c}
\hat{p}_x \\
\hat{p}_y \\
\end{array}\right]
,\label{Def:a}
\end{eqnarray}
where $w_0$ is the intensity-$1/e^2$ radius
 of the vacuum state $|\mathrm{vac}\rangle$ along the transverse mode.
$\hat{x}, \hat{y}$ are operatorized coordinates,
 and the momentum operators $\hat{p}_x, \hat{p}_y$ are defined
 so as to satisfy the canonical commutation relation
 $[\hat{x},\hat{p}_x]=[\hat{y},\hat{p}_y]=i\hbar$.
As with the longitudinal mode, the vacuum state is defined to be
 $\hat{a}_{x,y}|\mathrm{vac}\rangle=0$.
The Hermite-Gaussian (HG) mode of the principal numbers $n_x,n_y$ can be written as
 $|\mathrm{HG}_{n_x n_y}\rangle = \{(\hat{a}_x^\dagger)^{n_x}(\hat{a}_y^\dagger)^{n_y}/\!\!\sqrt{n_x!n_y!}\}|\mathrm{vac}\rangle$,
 while the Laguerre-Gaussian (LG) mode of the azimuthal quantum number $l$ and the magnetic quantum number $m$ is written as
 $|\mathrm{LG}_l^{m}\rangle= \{(\hat{a}_+^\dagger)^{l+m}(\hat{a}_-^\dagger)^{l-m}/\!\!\sqrt{(l+m)!(l-m)!}\}|\mathrm{vac}\rangle$,
 where the $\hat{a}_\pm $ are defined as
 $\hat{a}_\pm \equiv (\hat{a}_x \mp i\hat{a}_y)/\!\!\sqrt{2}$,
 and the quantum numbers take the values of
 $\{n_x,n_y\}=0,1,2,\cdots$,
 $l=0,1,2,\cdots$
 and $m=-l,\cdots,+l$ \
Thus, the HG and LG modes represent the number of states that can be occupied by the transverse mode.
The orbital angular momentum (OAM) operator
 $\hat{l}_z\equiv \hat{x}\hat{p}_y-\hat{y}\hat{p}_x$
 may be more familiar than annihilation operators given in Eq. (\ref{Def:a});
because the OAM can be written using annihilation operators as
 $\hat{l}_z=(\hat{a}_+^\dagger\hat{a}_+-\hat{a}_-^\dagger\hat{a}_-)\hbar/2$,
the oscillators in Eq. (\ref{Def:a})
 can be regarded as Schwinger's model oscillators for OAM \cite{Sakurai}.

In this paper, we show how to generate the transverse modes
 for the odd cat
 $|\!\!\uparrow\rangle\equiv \{|\mathrm{vac}\rangle+|\mathrm{coh}\rangle\}/\!\!\sqrt{2N_\uparrow}$,
 the even cat $|\!\!\downarrow\rangle\equiv \{|\mathrm{vac}\rangle-|\mathrm{coh}\rangle\}/\!\!\sqrt{2N_\downarrow}$,
 and for any arbitrary superposition of these:
\begin{eqnarray}
|\theta,\phi\rangle = \cos(\theta/2)|x-\rangle+e^{i\phi}\sin(\theta/2)|x+\rangle, \label{Eq:ket_qubit}
\end{eqnarray}
 where $|\mathrm{coh}\rangle$ is a coherent state for the transverse mode
 with small amplitude ($|\langle\hat{a}_x\rangle|\sim 1$),
 $N_{\uparrow,\downarrow}$ are normalization factors,
 and $|x\mp \rangle \equiv \{|\!\!\uparrow\rangle\pm|\!\!\downarrow\rangle\}/\!\!\sqrt{2}$
 is the pole state of a CV qubit.
As a proof-of-principle experiment,
 we generated six CV qubit states
 and observed their probability distributions in $\hat{X}$ and $\hat{P}$ space; these
measurements corresponded to quadrature amplitude homodyne detection.
As the measured distributions agreed closely with
 theoretical expectations under which perfect purity is assumed,
and their generation was deterministic,
we can can expect that CV quantum computing would function more ideally on the transverse mode than on the longitudinal mode.

The rest of this paper is organized as follows.
In Sec. \ref{Sec:Waveoptics},
 we briefly review how quantum dynamical behavior can be simulated
 with optical transverse modes.
In Sec. \ref{Sec:Qubit},
 we show how to create CV qubits and perform homodyne detection
 on optical transverse modes.
In Sec. \ref{Sec:Wigner},
 we develop Wigner functions and probability distributions for
 eight typical CV qubit states.
In Sec. \ref{Sec:Experiment},
 we report on the results of a proof-of-principle experiment.
In Sec. \ref{Sec:Applications},
 we propose three applications, namely
 ``adiabatic control of beam profile,"
 ``phase shift keying on the transverse mode,''
 and ``quantum cryptography using CV qubit states."
In Sec. \ref{Sec:Comment}, we comment on related research, and, finally,
In Sec. \ref{Sec:Conclusion},
 we provide a summary for this paper.

\section{Simulating quantum dynamics on optical transverse modes} \label{Sec:Waveoptics}
We will derive the formula of the beam radius of a Gaussian beam as an example
 of the analogies existing between quantum mechanics and paraxial optics.
In the case of coherent light in a longitudinal mode,
 the component of the electric field $E_x$ at position $(x,y,z)$ and time $t$ can be written as
\begin{equation}
E_x(x,y,z,t) = \Psi(x,y,z)E_0\exp\left[ik(z-ct)\right],\label{Def:wavefunction}
\end{equation}
 where $k$ is the wave number, $c$ is the speed of light,
 and $E_0$ is the field amplitude.
Using Dirac notation, we can represent the envelope $\Psi(x,y,z)$ as
\begin{eqnarray}
\Psi(x,y,ct') = \langle x,y|\hat{U}(t')|\psi\rangle, \label{Eq:Psi_xyz}
\end{eqnarray}
 where $t'\equiv z/c$,
 $\hat{U}(t')$ is defined as the time evolution operator
 and $|x,y\rangle$ is the simultaneous eigenstate of $\hat{x}$ and $\hat{y}$.
A unitary operator $\hat{U}(t')$
can be represented using a Hamiltonian $\hat{H}$ as
 $\hat{U}(t')=\exp[-it'\hat{H}/\hbar]$.
In order to obey the paraxial Helmholtz equation,
 $\hat{H}$ should be
\begin{eqnarray}
\hat{H}=\frac{\hat{p}_x^2+\hat{p}_y^2}{2m'}, \label{Eq:FreeHamiltonian}
\end{eqnarray}
 where $m'\equiv \hbar k/c$ \cite{Gloge1969}.
Thus, it is seen that the time evolution of the envelope $\Psi(x,y,z)$
 looks like the wavefunction of a free particle of mass $m'$.
The propagator for a free particle
 $K(x,x';t')\equiv\langle x,y'|\hat{U}(t')|x',y'\rangle$
 is given by \cite{Sakurai}
\begin{eqnarray}
K(x,x';t') &=& \sqrt{\frac{m'}{2\pi i t'\hbar}}\exp\left[\frac{im'(x-x')^2}{2t'\hbar }\right].
\end{eqnarray}
Because the vacuum state is defined as $\hat{a}_{x,y}|\mathrm{vac}\rangle=0$,
 its wavefunction
 $\psi_\mathrm{vac}(x,y)\equiv \langle x,y|\mathrm{vac}\rangle$ is Gaussian:
\begin{eqnarray}
\psi_\mathrm{vac}(x,y) = \sqrt{N_0}\exp\left[-\frac{x^2+y^2}{w_0^2}\right], \label{Eq:psi_vac}
\end{eqnarray}
 where $N_0$ is a normalization factor.
By integration of the product of the wavefunction and the propagator for the vacuum state,
 $\Psi(x,y,ct')=\int\!\!\int K(x,x';t')K(y,y';t')\psi(x',y')dx'dy'$,
 the general form of the Gaussian beam can be obtained:
\begin{eqnarray}
\Psi_\mathrm{vac}(x,y,z)=\sqrt{N_z}\exp\left[iP(z)+\frac{ik(x^2+y^2)}{2q(z)}\right],
\label{Def:GaussianBeam}
\end{eqnarray}
 where $N_z$ is a normalization factor,
 $P(z)\equiv -\mathrm{arctan}[z/z_R]$, and $z_R \equiv k w_0^2/2$.
 In paraxial wave optics, $q(z)\equiv 1/R(z)+2i/\{kw^2(z)\}$ is called the beam parameter,
 where $R(z)=z\{1+t(z_R/z)^2\}$ is the concave radius and
\begin{eqnarray}
w(z) = w_0\sqrt{1+(z/z_R)^2} \label{Eq:BeamDiameter}
\end{eqnarray}
 is the beam radius.
The derivation above is performed in detail in Appendix \ref{App:Paraxial}.

Using the same Hamiltonian (\ref{Eq:FreeHamiltonian}) as above, Eq. (\ref{Eq:BeamDiameter}) can also be derived from the same initial state (\ref{Eq:psi_vac})using Heisenberg matrices instead of the Schr\"{o}dinger equation.
 The Heisenberg equation $d\hat{A}/dt'=[\hat{A},\hat{H}]/(i\hbar)$
 gives the slope of the ray $\hat{v}_x\equiv d\hat{x}/dz=\hat{p}_x/(\hbar k)$.
Because the operator $\hat{A}$ evolves as
 $\hat{A}(t')\equiv \hat{U}^\dagger(t')\hat{A}\hat{U}(t')$,
 the time evolution of $\hat{x}$ and $\hat{v}_x$ can be summarized using the matrix
\begin{eqnarray}
\left[\begin{array}{c}
\hat{x}(t') \\
\hat{v}_x(t') \\
\end{array}\right]
&=& \left[\begin{array}{cc}
1 & z \\
0 & 1 \\
\end{array}\right]
\left[\begin{array}{c}
\hat{x} \\
\hat{v}_x \\
\end{array}\right]. \label{Eq:Ray-matrix}
\end{eqnarray}
This matrix is identical to the ray matrix in geometrical optics \cite{Yariv}.
$\hat{x}^2(t') = \hat{x}^2+ z (\hat{x}\hat{v}_x+\hat{x}\hat{v}_x)+z^2\hat{v}_x^2$
 obtained from Eq. (\ref{Eq:Ray-matrix}).
The vacuum state (\ref{Eq:psi_vac})
 has the distributions
 $\langle \hat{x}^2\rangle=w_0^2/4$ and
 $\langle \hat{v}_x^2\rangle=1/(kw_0)^2$.
According to Eq. (\ref{Def:GaussianBeam}),
 the beam radius is defined as $w(z)=2\sqrt{\langle\hat{x}^2(t')\rangle}$;
therefore, we can obtain the same answer as in Eq. (\ref{Eq:BeamDiameter}).
Note that the vacuum state satisfies the diffraction limit
 $(\Delta x)(\Delta v_x)=1/(2k)$,
 where we set $\Delta x\equiv\sqrt{\langle \hat{x}^2\rangle-\langle \hat{x}\rangle^2}$
 and $\Delta v_x\equiv\sqrt{\langle \hat{v}_x^2\rangle-\langle \hat{v}_x\rangle^2}$.

As shown above, the wavefunction of the vacuum state $\psi_\mathrm{vac}$
 is Gaussian in both the longitudinal and transverse modes.
On the other hand, the Hamiltonians of the respective modes differ; whereas the Hamiltonian for free propagation along the longitudinal mode
 is similar to that of a harmonic oscillator,
the Hamiltonian for free propagation along the transverse modes
 is similar to that of a free particle and can be written using Eq. (\ref{Eq:FreeHamiltonian}).
However, this difference is not critical for CV quantum computing, as the
Hamiltonian of a harmonic oscillator in transverse mode
 can be constructed using a single-lens system, as shown in Sec. \ref{Sec:Qubit}.

\section{Method for generating and observing a CV qubit} \label{Sec:Qubit}
\begin{figure}[htbp]
 \begin{center}
 	 \includegraphics[scale=0.4]{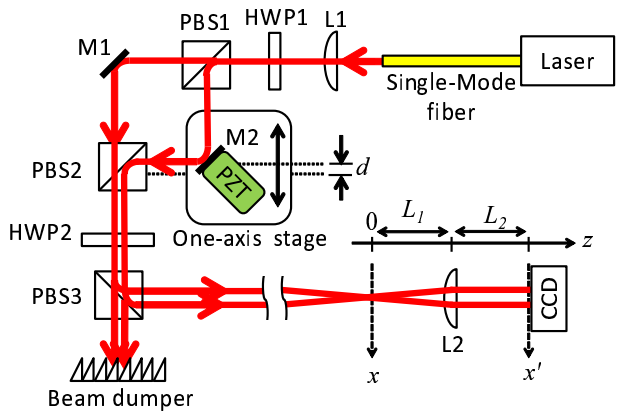}
 	 \caption{
 	 An apparatus for generating and observing a CV qubit on an optical transverse mode.
 	 A laser beam is spatially shaped with a single mode fiber and then split by
 	 half wave plate (HWP) 1 and polarizing beam splitter (PBS) 1 with transmittance $T$.
 	 Displacement $d$ and relative phase $\varphi$ are added between the two beams using
 	 mirrors (M) 1 and M2.
 	 By passing the beams through PBS2, HWP2, and PBS3, they are caused to interfere.
 	 Lens (L) 1 is set so that both beams are focused onto the $z=0$ plane.
 	 L2, with focal length$f$, is placed at $z = L_1$, and a CCD camera placed at $z=L_1+L_2$
 	 acquires the position or momentum distributions.
 }
 \label{fig:tabletop}
 \end{center}
\end{figure}

Fig. \ref{fig:tabletop} shows the process by which a CV qubit
is generated on an optical transverse mode.
A laser beam is passed through a single mode fiber in order to shape its transverse mode
 before being reflected by a mirror, (M) 1, passed through a polarizing beam splitter, (PBS) 3,
 and finally focused at $z=0$ by a lens, (L) 1.
The vacuum state $\psi_\mathrm{vac}(x,y)$ is generated at the focal plane ($z = 0$).
To generate a displaced vacuum state $\psi_\mathrm{vac}(x-d,y)$
 on the same focal plane,
 the beam can be split by a half wave plate, (HWP) 1, and by PBS1, with half of the beam then
 reflected by M2 and PBS2.
We can set $T$ ($0\le T \le 1$) as the transmittance at PBS1,
 $d$ as the misalignment from the vacuum state along the $x$-axis at M2,
 and $\varphi$ ($0\le \varphi < 2\pi$)
 as the relative phase between the beams passed through M1 and M2.
These values of $T$, $d$, and $\varphi$ can be adjusted using HWP1,
 a one-axis stage, and a piezo-electric transducer (PZT), respectively.
Because the polarizations of the beams passed through M1 and M2 are mutually orthogonal,
 HWP2 is inserted in order to cause them to interfere.
The unneeded beam from the other output port of PBS3 is
 absorbed by a beam dumper.
The field envelope formed at the focal plane by this configuration becomes
\begin{eqnarray}
\psi_\mathrm{arb}(x,y) =
&&\frac{1}{\sqrt{N_\mathrm{arb}}}\left\{\sqrt{T}\psi_\mathrm{vac}(x,y)\right.\nonumber\\
&&\left.+e^{i\varphi}\sqrt{1-T}\psi_\mathrm{vac}(x-d,y)\right\},
\label{Eq:psi_arb}
\end{eqnarray}
where $N_\mathrm{arb}$ is the normalization factor.

To check the equivalence between the ket representation in Eq. (\ref{Eq:ket_qubit})
 and the field envelope written as Eq. (\ref{Eq:psi_arb}),
 we derive the wavefunction for the coherent state
 $\psi_\mathrm{coh}(x,y) \equiv\langle x,y|\mathrm{coh}\rangle$.
As with the longitudinal mode,
 a coherent state on transverse mode is generated
 by using a displacement operator
 $\hat{D}_i(\alpha)\equiv \exp[\alpha\hat{a}_i^\dagger-\alpha^*\hat{a}_i]$
 on the vacuum state $|\mathrm{vac}\rangle$,
 i.e., $|\mathrm{coh}\rangle=\hat{D}_i(\alpha)|\mathrm{vac}\rangle$.
To simplify this formulation,
 we assume that the complex amplitude $\alpha$
 is a real number $\alpha=d_0/(\!\!\sqrt{2}w_0)$
 and that the direction of the displacement is along the $x$-axis;
using these assumptions, the displacement operator becomes $\hat{D}_x(\alpha)=\exp(-id\hat{p}_x/\hbar)$.
From the relation $\hat{D}_x^\dagger(-\alpha)\hat{x}\hat{D}_x(-\alpha)=\hat{x}-d$,
 we obtain $\langle x,y|\hat{D}_x^\dagger(-\alpha)=\langle x-d,y|$.
In addition, we find that $\hat{D}_x^\dagger(-\alpha)=\hat{D}_x(\alpha)$.
From these, we obtain $\psi_\mathrm{coh}(x,y)=\langle x,y|\hat{D}_x(\alpha)|\mathrm{vac}\rangle=\psi_\mathrm{vac}(x-d,y)$ as the wavefunction of the coherent state
(the same conclusion is derived in Ref. \cite{Nienhuis1993}).
The ket representation in Eq. (\ref{Eq:psi_arb}) then becomes
\begin{eqnarray}
|\mathrm{arb}\rangle = \frac{1}{\sqrt{N_\mathrm{arb}}}\left\{\sqrt{T}|\mathrm{vac}\rangle
+ e^{i\varphi}\sqrt{1-T}|\mathrm{coh}\rangle\right\}, \label{Def:ket_arb}
\end{eqnarray}
 which is mathematically equivalent to a CV qubit as given in Eq. (\ref{Eq:ket_qubit}).
The relation between $(T,\varphi)$ and $(\theta,\phi)$ are discussed further in Sec. \ref{Sec:Wigner}.
When the angle $\theta_d$ ($0\le \theta_d <\pi/2$) is defined as
 $\cos\theta_d\equiv \langle\mathrm{vac}|\mathrm{coh}\rangle$,
 we obtain $\cos\theta_d=\exp(-|\alpha|^2)$ and
 $N_\mathrm{arb}=1+2\sqrt{T(1-T)}\cos\theta_d\cos\varphi$.

To observe the momentum distribution,
 L2 can be inserted at $z=L_1$
 and a CCD camera set at $z=L_1+L_2$
 to measure the intensity distribution, as shown schematically in Fig. \ref{fig:tabletop}.
We assume that the spacings are constant, as $L_1 = L2\equiv f(1-\cos\theta_L)$, where $f$ is the focal length of L2, which is the definition of the rotation angle in the phase space $\theta_L$.
The position $\hat{x}$ and the slope of the ray $\hat{v}_x$
 on the CCD plane become
\begin{eqnarray}
\left[\begin{array}{c}
\hat{x}' \\
\hat{v}'_x \\
\end{array}\right]
=\left[\begin{array}{cc}
\cos\theta_L & f_0\sin\theta_L\\
-\sin\theta_L/f_0 & \cos\theta_L \\
\end{array}\right]
\left[\begin{array}{c}
\hat{x} \\
\hat{v}_x \\
\end{array}\right], \label{Eq:rotation}
\end{eqnarray}
 where $f_0\equiv f\sin\theta_L$ is the conversion factor
 between $v_x$ and $x$ \cite{Stoler1981,Lohmann1993}.
The CCD camera acquires the intensity distribution on the $z=L_1+L_2$ plane.
To simplify the calculation of the distribution,
 we reduce the $y$ dependence using $I(x')\equiv \int|\Psi(x',y,L_1+L_2)|^2dy$.
When the lengths are set to $L_1 = L_2 = f$,
 the rotation angle becomes $\theta_L = \pi/2$
 and the resulting intensity distribution, $I(x') = \int|\tilde{\psi}(p_x,p_y)|^2dp_y$, reflects the momentum distribution,
 where $x' = p_xf/(\hbar k)$ and $\tilde{\psi}(p_x,p_y)$ is the wavefunction in momentum space.
When $L_1 = L_2 = 0$, the intensity distribution reflects the position distribution 
 as $I(x) = \int|\psi(x,y)|^2dy$.

\section{Detailed description of CV qubit state} \label{Sec:Wigner}
We can define a reduced Wigner function of the $x$-mode as
\begin{eqnarray}
W(x,p_x) \equiv N_w\int\!\!\!\!\int_{-\infty}^{\infty}
&&\psi(x+x'/2,y)\psi^*(x-x'/2,y) \nonumber \\
&&\times\exp\left[-i\frac{x'p_x}{\hbar}\right]dx'dy, \label{Def:Wigner-p}
\end{eqnarray}
 where $N_w$ is a normalization factor that satisfies
 $\int\!\!\int W(x,p_x)dxdp_x=1$ \cite{Wodkiewicz1998}.
The Wigner function for the arbitrary superposition of coherent states given by Eq. (\ref{Eq:psi_arb}) then becomes
\begin{eqnarray}
&&W_\mathrm{arb}(x,p_x) =
\frac{1}{N_\mathrm{arb}}\left\{
TW_\mathrm{vac}(x,p_x)+(1-T)W_\mathrm{coh}(x,p_x)\nonumber\right. \\
&&\quad \left.+2\sqrt{T(1-T)}W_\mathrm{half}(x,p_x)\cos(\varphi-dp_x/\hbar)
\right\}, \label{Eq:wigner_arb}
\end{eqnarray}
 where $W_\mathrm{vac}(x,p_x)\equiv \exp[-2x^2/w_0^2-w_0^2p_x^2/(2\hbar^2)]/(\pi\hbar)$,
 $W_\mathrm{coh}(x,p_x) \equiv W_\mathrm{vac}(x-d,p_x)$, and $W_\mathrm{half}(x,p_x) \equiv W_\mathrm{vac}(x-d/2,p_x)$ are the Wigner functions for for the vacuum state, the coherent state with displacement $d$, and the coherent state with displacement $d/2$, respectively.

\begin{figure}[htb]
 \begin{center}
 \includegraphics[scale=0.65]{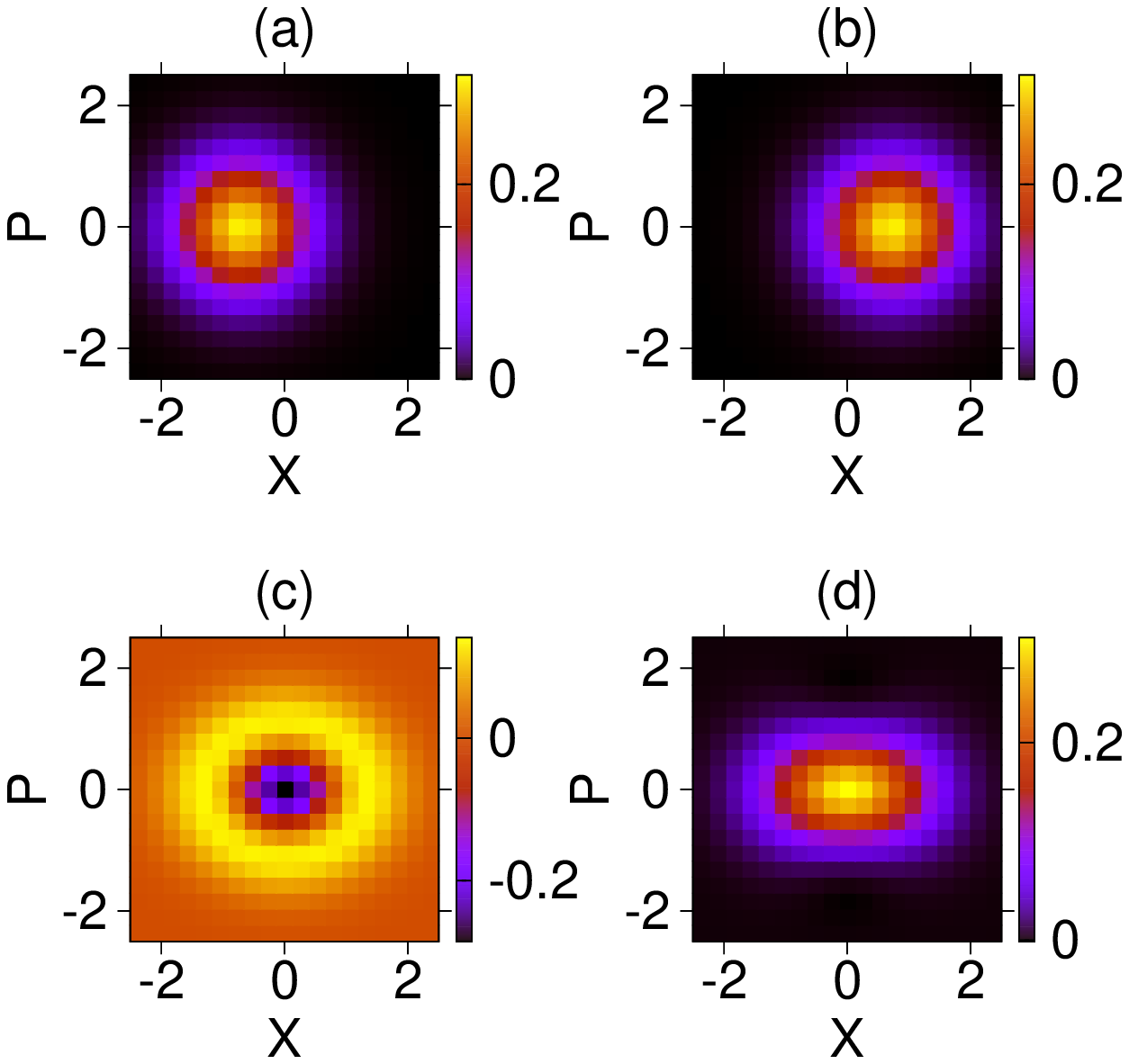}
 \includegraphics[scale=0.65]{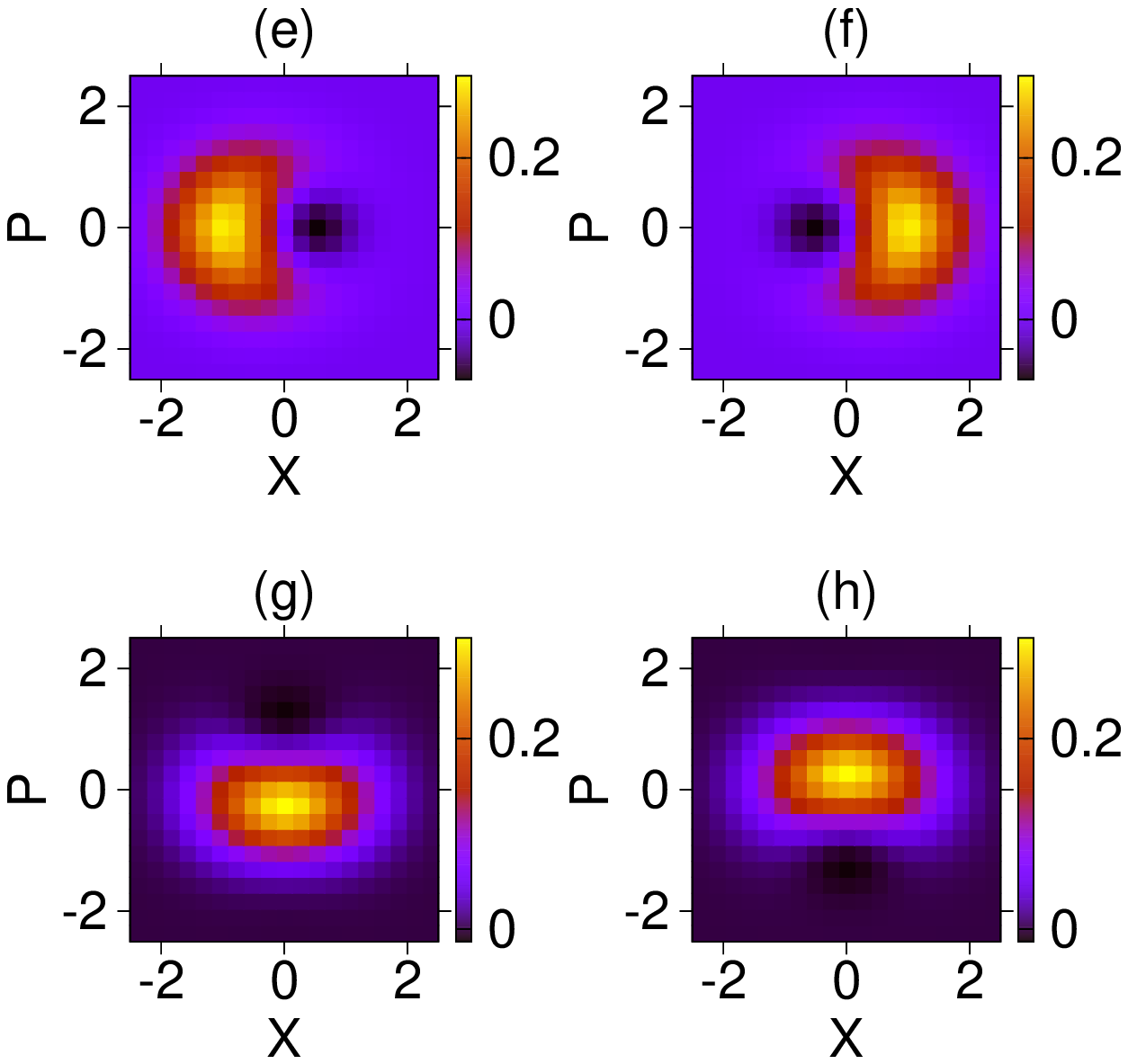}
 \caption{
 The Wigner functions for:
 (a) vacuum state $|\mathrm{vac}\rangle$,
 (b) coherent state $|\mathrm{coh}\rangle$,
 (c) odd cat state $|\!\downarrow\rangle$,
 (d) even cat state $|\!\uparrow\rangle$,
 (e) $|x-\rangle$,
 (f) $|x+\rangle$,
 (g) $|p_x-\rangle$,
 (h) $|p_x+\rangle$.
 This figure introduces the non-dimensional position and momentum,
 $X \equiv \sqrt{2}(x-d/2)/w_0$ and $P \equiv w_0 p_x/(\!\!\sqrt{2}\hbar)$,
 respectively.
 The displacement is set to $d=w_0$ ($|\alpha|=1/\sqrt{2}$).
 The position distributions of (e),(f) are squeezed,
 and the momentum distributions of (d),(g),(h) are squeezed.
 }
 \label{Fig:wigner-all}
 \end{center}
\end{figure}

To the already introduced typical states
$|\mathrm{vac}\rangle$,
$|\mathrm{coh}\rangle$,
$|\!\!\downarrow\rangle$,
$|\!\!\uparrow\rangle$,
and $|x\mp \rangle$,
we can introduce
$|p_x\mp \rangle\equiv \{|\!\!\uparrow \rangle \pm i|\!\!\downarrow \rangle\}/\sqrt{2}$.
The condition $T=1,0$ provides the vacuum and coherent states $|\mathrm{vac}\rangle$
 and $|\mathrm{coh}\rangle$, respectively.
The conditions $T=1/2$ and $\varphi=0,\pi$ provide
 the even and odd cats $|\!\!\uparrow\rangle$ and $|\!\!\downarrow\rangle$, respectively.
The condition $T=(1\pm \sin\theta_d)/2$ and $\varphi=\pi$ provides the
 $|x\mp\rangle$ state.
The condition $T=1/2$ and $\varphi=\mp\theta_d$ provides the
 $|p_x\mp\rangle$ state.
A detailed derivation of these states is provided in Appendix \ref{App:Typical}.
The Wigner functions at the above conditions are plotted in Fig. \ref{Fig:wigner-all},
 which introduces the non-dimensional position $X \equiv \sqrt{2}(x-d/2)/w_0$
 and the non-dimensional momentum $P \equiv w_0 p_x/(\!\!\sqrt{2}\hbar)$.
According to the quantization in Eq. (\ref{Def:a}),
these variables can be rewritten as
 $\hat{X}= (\hat{a}_x+\hat{a}_x^\dagger)/2$ and
 $\hat{P}= (\hat{a}_x-\hat{a}_x^\dagger)/(2i)$,
and their commutation relation becomes $[\hat{X},\hat{P}]=i/2$.
From Fig. \ref{Fig:wigner-all}(c) it is seen that
 $|\!\!\downarrow\rangle$ is similar to a one-number state, and from
Figs. \ref{Fig:wigner-all}(d)-(h) it is seen that
the $|x\mp\rangle$ are position squeezed states
 and that $|\!\!\uparrow\rangle$ and $|p\mp\rangle$ are momentum squeezed states.
The exact position and momentum distributions $I(x)$ and $\tilde{I}(p_x)$,respectively,
 can be calculated from the Wigner function as
 $I(x)=\int W(x,p_x)dp_x$ and $\tilde{I}(p_x)=\int W(x,p_x)dx$,
yielding
\begin{eqnarray}
I_\mathrm{arb}(x)&=&\frac{1}{N_\mathrm{arb}}
\left\{TI_\mathrm{vac}(x)+(1-T)I_\mathrm{coh}(x)\right.\nonumber \\
&&\left.+2\sqrt{T(1-T)}I_\mathrm{half}(x)\cos\theta_d\cos\varphi\right\}, \label{Eq:dist-pos}\\
\tilde{I}_\mathrm{arb}(p_x) &=& \frac{\tilde{I}_\mathrm{vac}(p_x)}{N_\mathrm{arb}}
\nonumber\\
&&\times \left\{1+2\sqrt{T(1-T)}\cos\left[\varphi-\frac{dp_x}{\hbar}\right]\right\}, \label{Eq:dist-mom}
\end{eqnarray}
 where we set
 $I_\mathrm{vac}(x)\equiv \int W_\mathrm{vac}(x,p_x)dp_x$,
 $I_\mathrm{coh}(x)\equiv I_\mathrm{vac}(x-d)$,
 $I_\mathrm{half}(x)\equiv I_\mathrm{vac}(x-d/2)$,
 and $\tilde{I}_\mathrm{vac}(p_x)\equiv \int W_\mathrm{vac}(x,p_x)dx$.
To derive the distributions
 in Eq. (\ref{Def:Wigner-p}), we assume that the purity of the CV qubit is unity; if this is not so, then $\psi(x+x'/2,y)\psi^*(x-x'/2,y)$ in Eq. (\ref{Def:Wigner-p})
 should be replaced by the matrix element of the density operator $\hat{\rho}$
 as $\langle x+x'/2,y~|~\rho~|~x-x'/2,y\rangle$.
Nevertheless, as we will show in Sec. \ref{Sec:Experiment},
 Eqs. (\ref{Eq:dist-pos}) and (\ref{Eq:dist-mom})
 agree closely with the experimental results.

\begin{figure}[htbp]
 \begin{center}
 \includegraphics[scale=0.30]{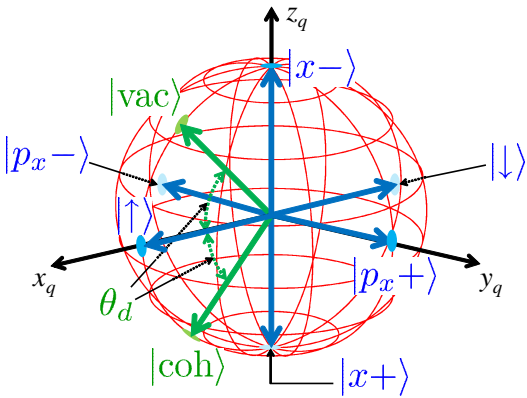}
 \caption{
 Bloch vector representation of the eight typical CV qubit states.
 Taking the large displacement limit ($\theta_d\to\pi/2$),
 the vacuum and coherent states approach the pole state as
 $|\mathrm{vac}\rangle \to |x-\rangle$ and
 $|\mathrm{coh}\rangle \to |x+\rangle$, respectively.
 }
 \label{fig:map-all}
 \end{center}
\end{figure}

To prepare a desired CV qubit state (\ref{Eq:ket_qubit})
 by arbitrary superposition of coherent states (\ref{Def:ket_arb}),
 $T$ and $\varphi$ must be set as
\begin{eqnarray}
T &=& \frac{1}{2}\left(1+\frac{z_q\sin\theta_d}{1-x_q\cos\theta_d}\right), \label{Eq:T}\\
\cos\varphi &=&
\frac{x_q-\cos\theta_d}{\sqrt{(x_q-\cos\theta_d)^2+(y_q\sin\theta_d)^2}},\label{Eq:cosvarphi}\\
\sin\varphi &=&
\frac{y_q\sin\theta_d}{\sqrt{(x_q-\cos\theta_d)^2+(y_q\sin\theta_d)^2}}, \label{Eq:sinvarphi}
\end{eqnarray}
 where $(x_q,y_q,z_q)\equiv (\sin\theta\cos\phi,\sin\theta\sin\phi,\cos\theta)$
 are the components of the Bloch vector of the qubit.
Fig. \ref{fig:map-all} shows the Bloch vectors for the eight typical qubit states;
a detailed derivation of these is given in Appendix \ref{App:Mapping}.
Although the Wigner functions for these states take many forms,
they can each be classified as a type of Bloch state
 with basis $\{|x-\rangle,|x+\rangle\}$.
Fig. \ref{fig:map-all} clearly shows that
 $\{|\mathrm{vac}\rangle, |\mathrm{coh}\rangle\}$
 is not strictly orthogonal as long as $\theta_d < \pi/2$,
 $\{|x-\rangle, |x+\rangle\}$ and $\{|p_x-\rangle, |p_x+\rangle\}$
 are orthogonal.
At the limit $\theta_d\to\pi/2$ ($d/w_0 \gg 1$),
 which corresponds to the orthogonal-state approximation between
 $|\mathrm{vac}\rangle$ and $|\mathrm{coh}\rangle$,
 $|\mathrm{vac}\rangle \to |x-\rangle$, and
 $|\mathrm{coh}\rangle \to |x+\rangle$.

\section{Experimental results from generation to observation of CV qubits} \label{Sec:Experiment}

As a proof-of-principle experiment,
 we generated six typical CV qubit states, namely
 $|\mathrm{vac}\rangle$, $|\mathrm{coh}\rangle$,
 and four states on the equator of the Bloch sphere.
We used a commercial external-cavity diode laser (toptica DL100) with
 wavelength $\lambda=2\pi/k=780$ nm to generate a beam that was transmitted through a single-mode fiber (780HP)
in order to shape its spatial distribution.
The power in front of PBS1 was 50 $\mu$W and
the visibility of the interferometer was measured to be 0.97 by setting $T=1/2$ and $d=0$.
A $f=145$ mm lens was used as L2.
The CCD camera for measuring the results output a monochromatic $720~\mathrm{pixel} \times 480~\mathrm{pixel}$ video signal
with an eight-bit analog-to-digital (A/D) resolution and a pixel size of 6.5 $\mu$m.
To reduce the background light, an iris and a neutral-density (ND) filter
 of optical density (OD) 2 was inserted in front of the CCD sensor.
To avoid saturating the video signal, an ND filter of OD3 and a combination of a HWP and a PBS
were inserted into the optical path, although such attenuation would be unnecessary if the exposure time of the CCD camera could be shortened.
The video signal output was acquired by a computer in order to
 analyze the distribution with $y$-dependence reduction, background removal and normalization.
Instead of measuring the position
 and momentum distributions with the same CCD camera,
 we inserted a non-polarizing beam splitter at $z<0$ in order to split each CV qubit beam into two beams for which we could view the respective distributions using two CCD cameras simultaneously.

\begin{figure}[htbp]
 \begin{center}
 \includegraphics[scale=0.5]{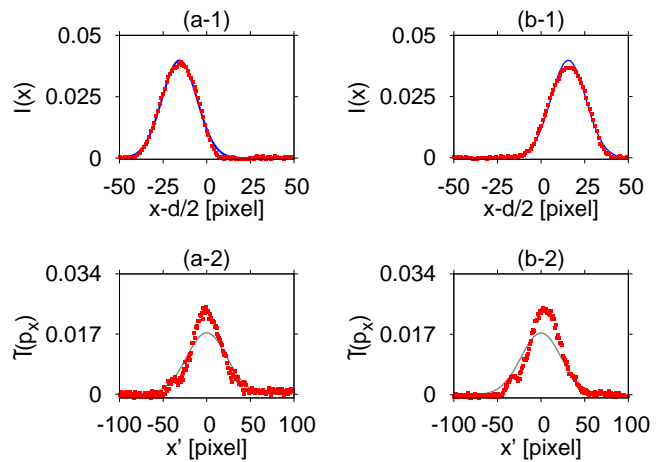}
 \caption{
 Experimental results of distribution measurement (red squares)
 (a-1,a-2) Position and momentum distributions, respectively, of the vacuum state.
 (b-1,b-2) Position and momentum distributions, respectively, of the coherent state.
 The position distributions can be fitted to Gaussian distributions (blue lines).
 The momentum distributions mostly obey theoretical predictions (gray lines).
 }
 \label{Fig:heterodyne-coh}
 \end{center}
\end{figure}

Fig. \ref{Fig:heterodyne-coh} shows the observed position and momentum distributions
 for the vacuum and the coherent states.
Fig. \ref{Fig:heterodyne-coh} (a-1) shows the position distribution of the vacuum state; a Gaussian fitting curve as given by Eq. (\ref{Eq:dist-pos}) with fitting parameter $w_0$
 is also shown.
Fig. \ref{Fig:heterodyne-coh} (b-1) shows the position distribution of the coherent state;
 again, a Gaussian fitting curve given by Eq. (\ref{Eq:dist-pos})
 with fitting parameter at the center $d$ is shown.
The above fittings produced $w_0=0.12$ mm ($z_R=60$ mm)
 and $\alpha=1.1$ ($\theta_d=0.40 \pi$).
Figs. \ref{Fig:heterodyne-coh} (a-2) and (b-2)
 show the momentum distributions of the vacuum and coherent states, respectively;
 the theoretical curves given by Eq. (\ref{Eq:dist-mom}) are also shown.
It is seen that these experimental results agree closely with theoretical predictions, with the exception of the half width maximum at $1/e^2$, which was experimentally determined to be 35 pixels, while the theoretical prediction was 46 pixels.
This difference might be caused by an axial displacement
 between the SM fiber and the aspherical lens inserted behind it,
 which would degrade the singularity of the spatial mode.
Another inconsistency was that, while theory predicted that Figs. \ref{Fig:heterodyne-coh} (a-2) and (b-2) would have the same center, the experimentally derived center in Fig. \ref{Fig:heterodyne-coh} (b-2) was five pixels further to the right.
One reason for this shift is that
 the PZT modified not only the relative phase $\varphi$
 but also the direction of $\langle\hat{v}_x\rangle$.
Replacing the PZT with an electro-optic modulator or a liquid crystal
 would be helpful in removing this error.
These differences notwithstanding,
 the similarity between Figs. \ref{Fig:heterodyne-coh} (a-2)
 and (b-2)
 indicates that the rotation angle $\theta_L$ is quite close to $\pi/2$,
 which in turn demonstrates that the momentum distributions have been successfully observed.

\begin{figure}[htbp]
 \begin{center}
 \includegraphics[scale=0.5]{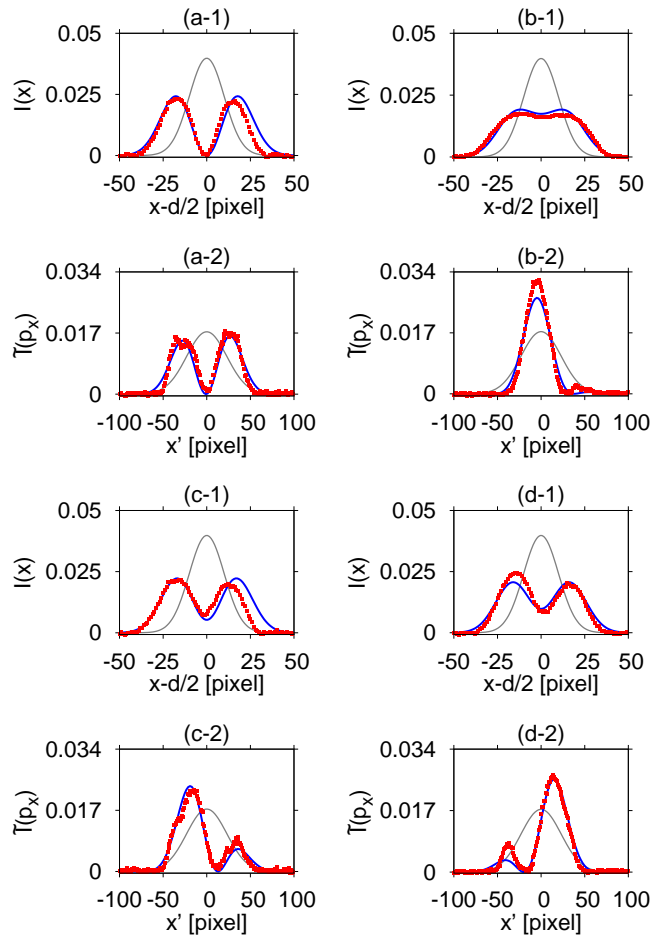}
 \caption{
 The position and momentum distributions of the four CV qubit states
 on the equator of the Bloch sphere (red squares).
 (a-1,a-2) Odd cat-like state. The relative phase is estimated to be $\varphi=0.98 \pi$.
 (b-1,b-2) Even cat-like state, with estimated relative phase of $\varphi=-0.18 \pi$.
 (c-1,c-2) $|p_x-\rangle$-like state, with estimated relative phase of $\varphi=-0.72\pi$.
 (d-1,d-2) $|p_x+\rangle$-like state, with estimated relative phase of $\varphi=0.57\pi$.
 The gray line indicates the standard quantum limit (SQL)
 and the blue lines are the theoretical curves based on estimation of $\varphi$.
 }
 \label{Fig:heterodyne-cat}
 \end{center}
\end{figure}

Fig. \ref{Fig:heterodyne-cat} shows the observed position and momentum distributions
 for the four CV qubit states, which
have been generated by fixing $T=1/2$ and setting four arbitrary values of $\varphi$.
The theoretical curves for $I_\mathrm{half}(x)$ or $\tilde{I}_\mathrm{half}(x)$
 are also plotted as a reference for the standard quantum limit (SQL).
At the relative phase obtained in Figs. \ref{Fig:heterodyne-cat} (a-1,a-2),
 the position and momentum distribution are both split into two peaks,
 which are characteristics of the odd cat state.
At the relative phases obtained in Figs. \ref{Fig:heterodyne-cat} (b-1,b-2)(c-1,c-2)(d-1,d-2),
 all of the position distributions are broader than the SQL,
 while all of the momentum distributions are narrower;
accordingly, these represent momentum squeezed states.
The peaks of the momentum distributions of Figs. \ref{Fig:heterodyne-cat} (b-2),
 (c-2),and (d-2)
 are close to the center, the negative side, and the positive side, respectively, of the distributions.
By fitting these momentum distributions with Eq. (\ref{Eq:dist-mom}),
 we were able to estimate the respective values of $\varphi$;
The fitted curves are shown as blue lines in Figs. \ref{Fig:heterodyne-cat} (a-2), (b-2),
 (c-2), and (d-2).
Based on these estimated values of $\varphi$ and Eq. (\ref{Eq:dist-pos}),
 we plotted the theoretical curves of the position distributions
 in Fig. \ref{Fig:heterodyne-cat} (a-1), (b-1), (c-1), and (d-1);
all of these agree closely with theoretical predictions.
The shapes of the Wigner functions will be
 similar to those of the $|\!\!\downarrow\rangle$, $|\!\!\uparrow\rangle$ and $|p_x \mp \rangle$ distributions.
To directly measure the relative phases $\varphi$ in Eq. (\ref{Def:ket_arb}) as well as the transmittance $T$, another beam not displaced by mirror M2 would be required;
then, by taking measurements at a number of values of $\theta_L$ in addition to $\theta_L=0,\pi/2$,
tomographic reconstruction of the Wigner function, as well as homodyne tomography on the quadrature amplitude, would be possible.
Although the image of the Wigner function $|W(x,p_x)|$ can be constructed by optical means \cite{Bartelt1980}, it is necessary to use
the tomographic method to observe the negative portion, $W(x,p_x)<0$,
which is needed to verify whether or not the state is non-Gaussian \cite{Dragoman2000}.

\section{Other Applications of CV qubit on transverse mode} \label{Sec:Applications}
As we explained in Sec. \ref{Sec:Intro}, CV qubits on transverse modes are useful in achieving CV quantum computing.
In this section, we will propose three further applications of CV qubits on transverse modes.

The first of these is ``adiabatic control of the beam profile.''
As shown in Fig. \ref{Fig:wigner-all},
 many varieties of non-Gaussian state can be produced
 on the same focal plane by choosing values for the two experimental parameters
 $(T,\varphi)$.
In particular, we can examine
 the intensity distributions on the focal plane $\Delta x$
 and the mean values of the slope of ray $\langle \hat{v}_x\rangle$.
It is seen that $\Delta x$ of $|x\mp\rangle$ is narrower than the SQL and that
the $\langle \hat{v}_x\rangle$ of $|p_x\mp\rangle$
 has either a positive or a negative value.
The intensity distribution of $|\!\!\downarrow\rangle$
 has a local minimum at its center.
Thus, by continuously varying $T$ and $\varphi$,
 the spatial distribution $\Delta x$ and the slope of ray $\langle \hat{v}_x\rangle$
 can be continuously modified without replacing or mechanically tilting optical elements.
This would be useful in the optical dipole trapping of cold atoms or
 as an optical tweezer for a microscope.
Note that a similar phenomenon was predicted
 in Ref. \cite{Steuernagel2005AJP}
 and demonstrated in Ref. \cite{Steuernagel2005JMO}.
In these methods, interference between the two higher order HG modes
 $|\mathrm{HG}_{40}\rangle$ and $|\mathrm{HG}_{50}\rangle$ were utilized.
By contrast, our method requires the two lowest HG modes
 $|\mathrm{vac}\rangle$ and $|\mathrm{coh}\rangle$.
More ideal beams for such applications can be generated experimentally.

A second application for transverse mode CV cubits is
 ``phase shift keying (PSK) on transverse modes.''
 Multiplexing light with differing spatial modes by applying phase shift keying (PSK) on the quadratic amplitude, or mode division multiplexing (MDM), has gained interest as a means for increasing capacity in optical fiber communications \cite{Amin2011};
Typically, four transverse mode states, such as
 $\{|\mathrm{HG}_{00}\rangle$,
 $|\mathrm{HG}_{10}\rangle$,
 $|\mathrm{HG}_{01}\rangle$,
 $|\mathrm{HG}_{11}\rangle\}$
 or $\{|\mathrm{LG}_0^{(0)}\rangle$,
 $|\mathrm{LG}_{1}^{(-1)}\rangle$,
 $|\mathrm{LG}_{1}^{(0)}\rangle$,
 $|\mathrm{LG}_{1}^{(+1)}\rangle\}$
 are used as the bases of a four-MDM system \cite{Bozinovic2013}.
The cat states $\{|\!\!\uparrow\rangle, |\!\!\downarrow\rangle, |\!\!\uparrow_y\rangle, |\!\!\downarrow_y\rangle\}$
 can be also adopted as the four bases of an MDM,
 where we define
 $|\!\!\uparrow_y\rangle\equiv \{\hat{D}_y(-\alpha/2)+\hat{D}_y(\alpha/2)\}|\mathrm{half}\rangle/\!\!\sqrt{2N_\uparrow}$ and
 $|\!\!\downarrow_y\rangle\equiv \{\hat{D}_y(-\alpha/2)-\hat{D}_y(\alpha/2)\}|\mathrm{half}\rangle/\!\!\sqrt{2N_\downarrow}$.
Because these bases are mutually orthogonal,
 performance that is similar to that of a standard four-MDM system will be obtained.
These bases can be also be regarded as two sets of two transverse mode PSKs, i.e.,
 $\{|\!\!\uparrow\rangle, |\!\!\downarrow\rangle\}$ and
 $\{|\!\!\uparrow_y\rangle, |\!\!\downarrow_y\rangle\}$.
The number of bases can be increased by reducing the pitch of PSK for a transverse mode on the Bloch sphere;
for example, 
 $\{|\!\!\uparrow\rangle, |\!\!\downarrow\rangle, |p_x-\rangle, |p_x+\rangle, |x-\rangle, |x+\rangle\}$
 and
 $\{|\!\!\uparrow_y\rangle, |\!\!\downarrow_y\rangle, |p_y-\rangle, |p_y+\rangle, |y-\rangle, |y+\rangle\}$
 can be adopted as the twelve bases of an MDM,
 where we define
 $|y\mp \rangle\equiv (|\!\!\uparrow_y\rangle\pm |\!\!\downarrow_y\rangle)/(\!\!\sqrt{2})$ and
 $|p_y\mp \rangle\equiv (|\!\!\uparrow_y\rangle\pm i|\!\!\downarrow_y\rangle)/(\!\!\sqrt{2})$.
As a tradeoff for increasing the number of bases using this method,
the bit error rate on decoding will increase; however, the resulting system could still represent a more efficient method of increasing communication capacity
 than the standard MDM.
In standard MDM, 16 MDM is necessary to construct twelve bases,
 which means that $n_{x,y} \le 3$ of HG mode or $l\le 3$ of LG mode are required.
Because of the diffraction limit,
 spatial modes of higher order require a larger cross section;
as a result, transmission becomes more lossy
as the order of the spatial modes increases\cite{Li2012}.
Thus, it might be better overall to use PSK on two transverse mode than to use MDM.

The final application is ``quantum cryptography using CV qubit states.''
In standard quantum cryptography with CV,
 four PSK states for the quadrature amplitude
 $\{|-\alpha_\omega\rangle, |+\alpha_\omega\rangle, |-i\alpha_\omega\rangle, |+i\alpha_\omega\rangle\}$, where we define $|\beta_\omega\rangle\equiv \hat{D}_\omega(\beta)|0_\omega\rangle$
 and $|0_\omega\rangle$ is the vacuum state of both the longitudinal and transverse modes,
are sent randomly \cite{Hirano2003}.
In quantum cryptography with CV qubit states, by contrast,
the qubit states $\{|x-\rangle, |x+\rangle, |p_x-\rangle, |p_x+\rangle\}$
 are sent randomly.
Because two pairs of these are mutually orthogonal, $\langle x-|x+\rangle=\langle p_x-| p_x+\rangle=0$,
 and the same level of security as in the Bennett-Brassard 1984 (BB84) protocol
 with weak coherent light can be obtained.
In a Graded - Index (GI) fiber,
 propagation through a phase shifter can be described using
 $\hat{H}=\hbar \omega_\mathrm{GI}( \hat{a}_x^\dagger \hat{a}_x+\hat{a}_y^\dagger \hat{a}_y)$
 instead of Eq. (\ref{Eq:FreeHamiltonian}).
The zigzag period of rays within a GI fiber, $cT'\equiv 2\pi c/\omega_\mathrm{GI}$,
 is typically on the order of 1 mm, which is much larger than the wavelength $cT' \gg \lambda$.
Thus, discriminating the four qubit states $\{|x-\rangle, |x+\rangle, |p_x-\rangle, |p_x+\rangle\}$
 is possible unless the optical path length fluctuates on the order of $cT'$.
This limitation is very loose compared with that
 associated with standard quantum cryptography using CV.
Discriminating the four longitudinal coherent states
 $\{|-\alpha_\omega\rangle, |+\alpha_\omega\rangle, |-i\alpha_\omega\rangle, |+i\alpha_\omega\rangle\}$
 is possible unless the optical path length fluctuates on the order of $\lambda$.
Note that, whereas qubit states in the longitudinal mode are easily decohered by transmission losses,
those in transverse mode are conserved after transmission losses.
Thus, quantum cryptography with CV qubit states is a good example of an application that is
 difficult in the longitudinal mode but easy in the transverse mode.

\section{Comments on relevant research} \label{Sec:Comment}

Much relevant work based on coherent light has been conducted by other researchers.
Discrete-variable quantum computing with numbered states on two transverse modes,
 such as the LG and HG modes, is demonstrated in Ref. \cite{Oliveira2005}.
CV quantum computing
 using optical fiber is discussed in Ref. \cite{Man'ko2001}.
The even cat for a transverse mode was generated
 in the large displacement regime ($|\alpha|^2\gg 1$),
 and the absolute value of the Wigner function $|W(x,p_x)|$
 was observed by optical means in Ref. \cite{Dragoman2001}.
 Ref. \cite{Dragoman2001}
 cautioned that the
 Wigner function for the field envelope simply mimics
 the form of the Wigner function in quantum mechanics.
One reason for this is that no decoherence was naturally induced in their study;
in our system, by contrast, there is a decoherence mechanism.
According to Eq. (\ref{Eq:wigner_arb}),
 the mixed state
 $W_\mathrm{mix}=(W_\mathrm{vac}+W_\mathrm{coh})/(2N_\mathrm{arb})$
 can be obtained by setting $T=1/2$ and averaging $\varphi$.

The beam splitter is also one of the most important elements in CV quantum computing,
and in such applications, beam splitter functionality similar to that for the longitudinal mode can be obtained for the transverse mode by using a Kerr nonlinear medium, as shown, for example, in
 \cite{Chavez-Cerda2007,Mar-Sarao2008}.

Another class of relevant research has focused on few-photon states
 and the additional spatial degrees of freedom that can be utilized in these.
Two-photon states with a large transverse displacement regime, called the spatial qubit,
have been generated by
 \cite{Lima2006,Taguchi2008},
and entangling the transverse modes of a few-photon state
 is considered as a resource for CV quantum computing in \cite{Tasca2011, Avelar2013}.

\section{Summary} \label{Sec:Conclusion}

Based on the quantization of optical transverse modes defined in Eq. (\ref{Def:a})
 and in Ref. \cite{Nienhuis1993},
we experimentally generated continuous-variable (CV) qubits on an optical transverse mode.
A CV qubit, which is a class of non-Gaussian state
 defined by the superposition of two coherent states,
 is useful as an initial state preparation in CV quantum computing. The Wigner functions and the Bloch-vector representations of typical eight CV qubits were derived.
Arbitrary CV qubit states can be generated
using the experimental setup shown in Fig. \ref{fig:tabletop},
and by experimentally measuring the position and momentum distributions,
we verified the successful generation of Schr\"{o}dinger's cat
 and momentum-squeezed states.
This system is more robust, efficient, and practical than
those using standard CV qubits based on the optical longitudinal mode.
As further applications of CV qubits on transverse modes,
 we proposed ``adiabatic control of beam profile,''
 ``phase shift keying on transverse modes,''
 and ``quantum cryptography with CV qubit states.''

Finally, we must stress that
 these results represent only the first step in achieving
CV quantum computing with optical transverse modes.
As long as the requirements of CV quantum computing are satisfied,
 we do not need to judge whether or not the system studied here is a quantum system.
Needless to say, it is a macroscopic system consisting of coherent light in longitudinal mode.
Because the probability distribution is squeezed from the standard quantum limit (SQL)
 and the Wigner function of the odd cat state can assume negative values,
 it would be natural to regard this as a quantum system in terms of the transverse mode.
Nevertheless, even without regarding it as a quantum system,
 this system remains useful for CV quantum computing
 because it can produce a complete set of tools needed for CV computing.
This example, therefore, may lead to new interpretations of the nature of quantum objects or quantum computing.

\acknowledgments

This work is supported by MATSUO FOUNDATION and JSPS KAKENHI Grant Number 22340113.
We greatfully acknowledge the technical assistance of Hiroumi Toyohama
 in developing the method 
 to measure the intensity distribution of a laser beam. 

\appendix

\section{Hamiltonian identical to paraxial optics} \label{App:Paraxial}

We derive the Hamiltonian for free propagation given by Eq. (\ref{Eq:FreeHamiltonian}).
The wave equation is written as
\begin{eqnarray}
\left\{\frac{\partial^2}{\partial x^2}+\frac{\partial^2}{\partial y^2}
+\frac{\partial^2}{\partial z^2}-\frac{1}{c^2}\frac{\partial^2}{\partial t^2}\right\}E_x=0.
\end{eqnarray}
By using the envelope $\Psi(x,y,z)$ defined in Eq. (\ref{Def:wavefunction}),
 the partial derivative with $z$ can be written as
\begin{eqnarray}
\frac{\partial^2E_x}{\partial z^2}
= \left\{-k^2+2ik\frac{\partial \Psi}{\partial z}+\left(\frac{\partial\Psi}{\partial z}\right)^2+\frac{\partial^2\Psi}{\partial z^2}\right\}E_x.
\end{eqnarray}
By applying the slowly varying envelope approximation, the wave equation becomes the
paraxial Helmholtz equation
\begin{eqnarray}
\frac{\partial}{\partial z} \Psi = -\frac{1}{2ik}\left[\frac{\partial^2}{\partial x^2}+\frac{\partial^2}{\partial y^2}\right]\Psi. \label{Eq:Helmholtz}
\end{eqnarray}
According to Eq. (\ref{Eq:Psi_xyz}), the envelope evolves as
\begin{eqnarray}
\frac{\partial \Psi}{\partial t'} = \frac{1}{i\hbar}\langle x,y |\hat{H}\hat{U}(t')|\psi\rangle,
\label{Eq:Schrodinger}
\end{eqnarray}
 which is otherwise known as the Schr\"{o}dinger equation.
The momentum operator $\hat{p}_x$ acts on the position eigenstates as \cite{Sakurai}
\begin{eqnarray}
\langle x',y'|\hat{p}_x=\frac{\hbar}{i}\frac{\partial}{\partial x'}\langle x',y'|.
\label{Def:momentum}
\end{eqnarray}
Therefore, assuming $\hat{H}=(\hat{p}_x^2+\hat{p}_y^2)/(2m')$ and $m'\equiv \hbar k/c$,
the paraxial Helmholtz equation (\ref{Eq:Helmholtz}) and
the Schr\"{o}dinger equation (\ref{Eq:Schrodinger}) become identical.
The wavefunction of the vacuum state written as Eq. (\ref{Eq:psi_vac})
 is obtained from Eq. (\ref{Def:momentum}).
The Gaussian beam having the form of Eq. (\ref{Def:GaussianBeam}) is obtained
 by using the formula
\begin{eqnarray}
\int_{-\infty}^\infty e^{-(ax^2+bx+c)}dx = \sqrt{\frac{\pi}{a}}\exp\left[\frac{b^2-4ac}{4a}
\right]. \label{Eq:IntExp3}
\end{eqnarray}

\section{Generation of typical states} \label{App:Typical}

The normalization factors of the even $|\!\!\uparrow\rangle$ and odd cats $|\!\!\downarrow\rangle$
 introduced in Sec. \ref{Sec:Intro} are written as $N_\uparrow = 1+\cos\theta_d$ and
 $N_\downarrow = 1-\cos\theta_d$, respectively.
The typical states $|x\mp\rangle$ and $|p\mp\rangle$ are expanded
 by the vacuum state and coherent states as
\begin{eqnarray}
|x\mp\rangle &=&
\frac{1}{\sin\theta_d}\left[
\frac{c_d \pm s_d}{\sqrt{2}}|\mathrm{vac}\rangle+\frac{c_d\mp s_d}{\sqrt{2}}|\mathrm{coh}\rangle\right], \\
|p\mp\rangle &=&
\frac{1}{\sqrt{2}}\left[
\frac{c_d \pm i s_d}{\sin\theta_d}|\mathrm{vac}\rangle+\frac{c_d\mp is_d}{\sin\theta_d}|\mathrm{coh}\rangle\right],
\end{eqnarray}
 where we set $c_d\equiv \cos(\theta_d/2)=\sqrt{N_\uparrow/2}$ and
 $s_d\equiv \sin(\theta_d/2)=\sqrt{N_\downarrow/2}$ for simplicity.
Note that $\sin\theta_d=2c_ds_d=\sqrt{N_\uparrow N_\downarrow}$.
From the relations of $c_d \pm s_d = \sqrt{1\pm \sin\theta_d}$ and
$(c_d \mp is_d)^2 = \exp(\mp i\theta_d)$,
we obtain the condition for generating the typical states
 written in Sec. \ref{Sec:Wigner} as well.

\section{Generation of arbitrary states} \label{App:Mapping}

The expansion coefficients of the Bloch state given in Eq.(\ref{Eq:ket_qubit})
 for the vacuum and coherent states
 $C_\mathrm{vac}\equiv \langle\mathrm{vac}|\theta,\phi\rangle$ and
 $C_\mathrm{coh}\equiv \langle\mathrm{coh}|\theta,\phi\rangle$, respectively,
 become as follows:
\begin{eqnarray}
C_\mathrm{vac} &=& \frac{c_d+s_d}{\sqrt{2}\sin\theta_d}\cos\frac{\theta}{2}
-\frac{c_d-s_d}{\sqrt{2}\sin\theta_d}e^{i\phi}\sin\frac{\theta}{2},\\
C_\mathrm{coh} &=& -\frac{c_d-s_d}{\sqrt{2}\sin\theta_d}\cos\frac{\theta}{2}
+\frac{c_d+s_d}{\sqrt{2}\sin\theta_d}e^{i\phi}\sin\frac{\theta}{2}.
\end{eqnarray}
Then, we obtain
\begin{eqnarray}
|C_\mathrm{vac}|^2 &=&
\frac{1}{2\sin^2\theta_d}\left[1-x_q\cos\theta_d+z_q \sin\theta_d\right],\\
|C_\mathrm{coh}|^2 &=&
\frac{1}{2\sin^2\theta_d}\left[1-x_q\cos\theta_d-z_q \sin\theta_d\right],\\
C_\mathrm{coh} C_\mathrm{vac}^* &=& \frac{1}{2\sin^2\theta_d}\left[x_q-\cos\theta_d+iy_q\sin\theta_d\right].
\end{eqnarray}
The normalization factor $N_\mathrm{arb}$ becomes
\begin{eqnarray}
N_\mathrm{arb} &=& \frac{1}{|C_\mathrm{vac}|^2+|C_\mathrm{coh}|^2}=\frac{\sin^2\theta_d}{1-x_q\cos\theta_d}.
\end{eqnarray}
From the relation
$T=N_\mathrm{arb}|C_\mathrm{vac}|^2$ and
$\varphi=\mathrm{arg}[C_\mathrm{coh}C_\mathrm{vac}^*]$,
Eqs. (\ref{Eq:T})(\ref{Eq:cosvarphi})(\ref{Eq:sinvarphi}) are obtained.
By comparing the condition for generating the typical states derived in Appendix \ref{App:Typical},
 we can find a representation of the typical states with the Bloch vector $(x_q,y_q,z_q)$ shown in
Fig. \ref{fig:map-all}.

\bibliography{biblio}

\end{document}